\documentclass[compactaffiliation]{Interspeech}

\interspeechcameraready

\title{AuralNet: Hierarchical Attention-based 3D Binaural Localization of Overlapping Speakers}

\author[affiliation={1}, equalcontribution]{Linya}{Fu}
\author[affiliation={1}, equalcontribution]{Yu}{Liu}
\author[affiliation={1}]{Zhijie}{Liu}
\author[affiliation={1}]{Zedong}{Yang}
\author[affiliation={2}]{Zhong-Qiu}{Wang}
\author[affiliation={3}]{Youfu}{Li}
\author[affiliation={1}]{He}{Kong}

\affiliation{School of Automation and Intelligent Manufacturing}{Southern University of Science and Technology (SUSTech)}{China}
\affiliation{Department of Computer Science and Engineering}{SUSTech}{China}
\affiliation{Department of Mechanical Engineering}{City University of Hong Kong}{China}

\email{[12232297;12250044;12332642;12210810]@mail.sustech.edu.cn, wang.zhongqiu41@gmail.com, meyfli@cityu.edu.hk, kongh@sustech.edu.cn}


\usepackage{comment}
\usepackage{multirow}
\usepackage{booktabs} 

\begin{document}

\maketitle

\begin{abstract}
    
    We propose AuralNet, a novel 3D multi-source binaural sound source localization approach that localizes overlapping sources in both azimuth and elevation without prior knowledge of the number of sources. AuralNet employs a gated coarse-to-fine architecture, combining a coarse classification stage with a fine-grained regression stage, allowing for flexible spatial resolution through sector partitioning. The model incorporates a multi-head self-attention mechanism to capture spatial cues in binaural signals, enhancing robustness in noisy-reverberant environments. A masked multi-task loss function is designed to jointly optimize sound detection, azimuth, and elevation estimation. Extensive experiments in noisy-reverberant conditions demonstrate the superiority of AuralNet over recent methods. 
\end{abstract}

\section{Introduction}
Sound source localization (SSL) plays a pivotal role in applications such as acoustic scene analysis \cite{taherian2024,qian2021multi,tan2021deep}, robot audition \cite{rascon2017localization,su2021necessary,wang2024slam,fu2024asm}, and sound source tracking \cite{qian2024glmb,evers2020locata}. Traditional SSL methods, which rely on parametric models for direction-of-arrival (DOA) estimation, often underperform in complex environments characterized by noise, reverberation, and overlapping sources \cite{pang2017binaural, wu2015binaural}.
In contrast, the human auditory system excels in challenging \textit{cocktail party} scenarios \cite{haykin2005cocktail}, capable of robustly isolating and localizing target sounds. This remarkable capability has spurred recent advances in binaural SSL (BSSL) based on deep neural networks (DNNs) \cite{grumiaux2022survey,wang2018robust,wang2018robustTDOA}.

Despite these developments, most deep learning-based BSSL approaches focus on single-source or 2D localization, typically estimating only azimuth \cite{beit2020importance, fejgin2024comparison, phokhinanan2023binaural, karthik2018subband}.
Many methods only consider the frontal hemisphere or assume prior knowledge of the number of active sources \cite{hu2023, phokhinanan2024, vecchiotti2019, ding2020joint, ma2017}, 
thereby limiting their real-world applicability. 
The recent work in \cite{geva2024binaural} tackles the 3D multi-source BSSL problem, but does not explicitly address the effects of real-world noise and reverberation, which are inherent in practical environments. 
Some studies have explored sector-based approaches to mitigate source association ambiguity \cite{yang2024deepear, krause2021joint}. However, the spatial partition in \cite{krause2021joint} is rather coarse-grained.
In \cite{yang2024deepear}, a classification-then-regression strategy was proposed to improve azimuth estimation, but it lacks an elevation estimation module, making it unsuitable for 3D localization.

\begin{figure}[t]
\begin{minipage}[b]{0.47\linewidth}
  \centering
  \centerline{\includegraphics[width=0.9\linewidth]{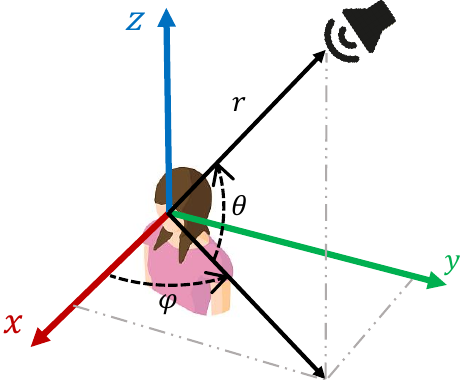}}
  \centerline{(a)}\medskip
\end{minipage}
\hfill
\begin{minipage}[b]{0.52\linewidth}
  \centering
  \centerline{\includegraphics[width=1\linewidth]{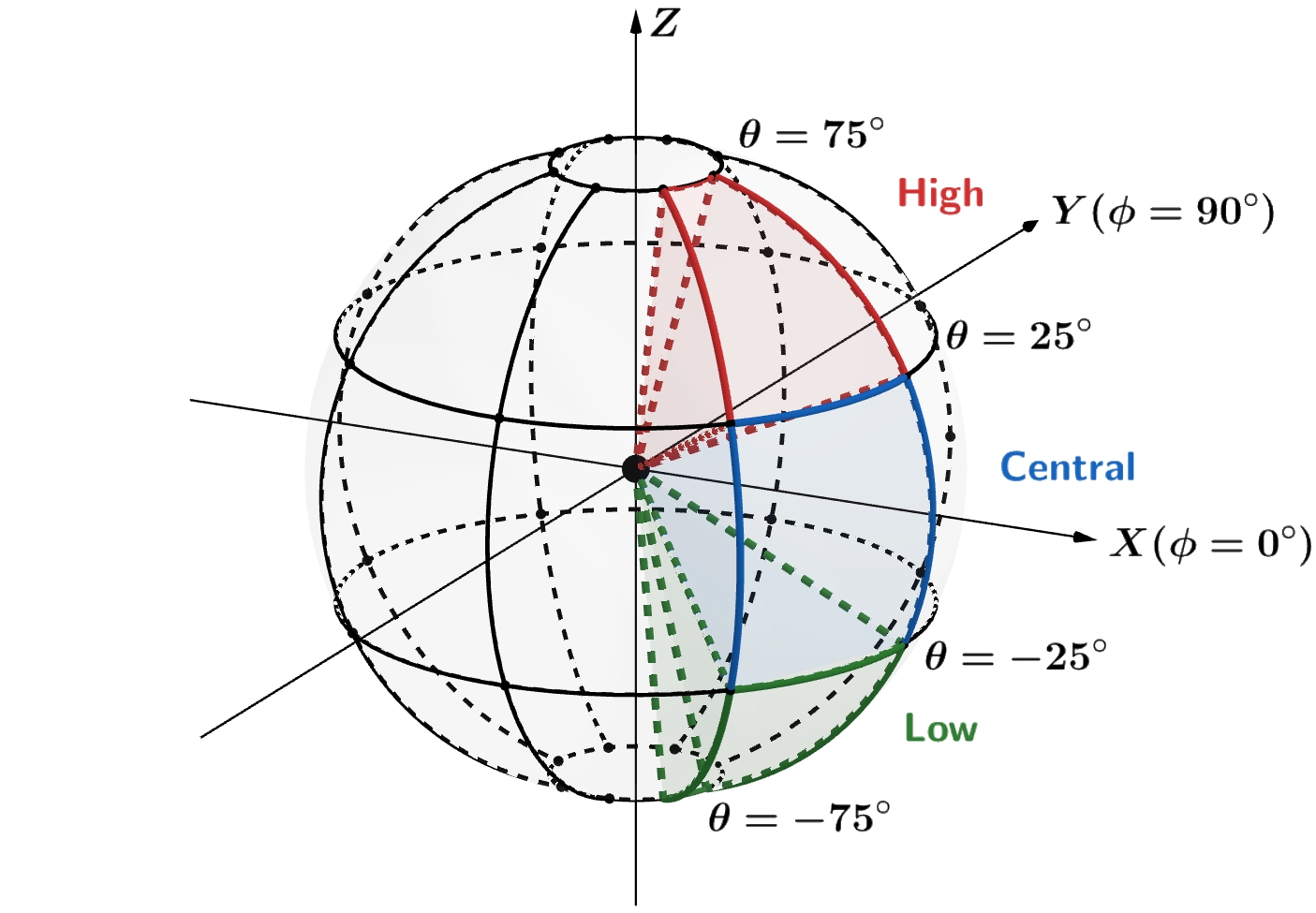}}
  \centerline{(b)}\medskip
\end{minipage}
\vspace{-7.5mm}
\caption{Illustration of (a) DOA representation in 3D; and (b) an example of dividing spherical sectors in a unit sphere with $M = 8$ and $N = 3$.}
\label{fig:sectors}
\vspace{-5mm}
\end{figure}

To address these limitations, we propose AuralNet, a novel hierarchical 3D multi-source BSSL framework that jointly estimates azimuth and elevation without requiring prior knowledge of the number of sound sources. Inspired by hierarchical processing observed in the human auditory system \cite{zhang2021hierarchical, fontolan2014contribution}, AuralNet employs a \textit{coarse-to-fine} strategy, where a coarse classification stage first identifies spatial azimuth sectors, followed by a fine-grained stage that refines elevation estimation within these sectors. This is complemented by two regression tasks that predict azimuth and elevation angles. We also integrate a self-attention mechanism from the Transformer architecture \cite{Vaswani2017attention} to enhance binaural feature fusion, mimicking the auditory cortex’s ability to focus on salient temporal cues in complex acoustic environments \cite{zatorre2001spectral, herdener2013spatial}.
In addition, we design a weighted loss function to coordinate the coarse and fine tasks, jointly optimizing source detection and angular estimation.

\begin{figure*}[t]
    \centering
    \includegraphics[width=1.0\textwidth]{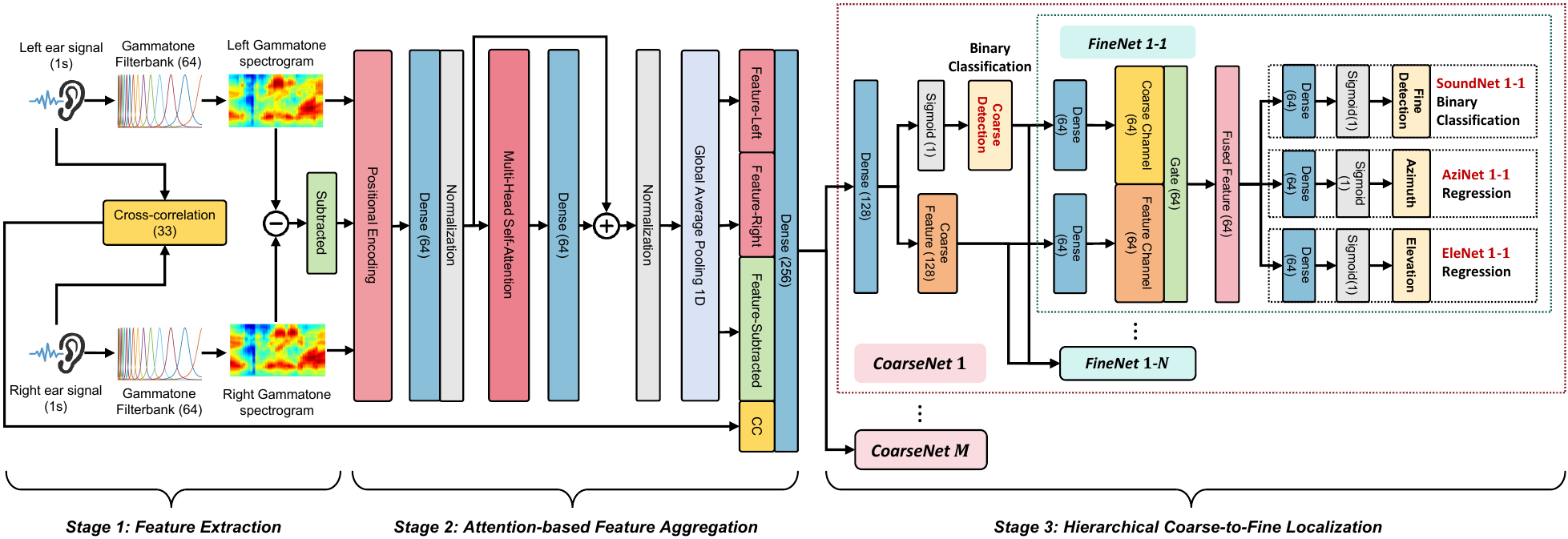}
    \vspace{-4mm}
    \caption{Architecture of proposed AuralNet.}
    \label{fig:network}
\vspace{-4mm}
\end{figure*}

The main contributions of this work are: 
(a) a robust 3D multi-source BSSL system capable of full $360^\circ$ azimuth and elevation localization, without prior knowledge of the number of sources; 
(b) a hierarchical coarse-to-fine multi-task learning framework facilitated by a gating mechanism that efficiently reduces model redundancy, enhances localization accuracy, and allows configurable sector partitioning for flexible spatial resolution;
(c) a multi-head self-attention mechanism to capture critical spatial cues from binaural features, improving robustness in noisy environments by modeling spectrotemporal dependencies and refining disparity features across the left and right ear spectra; and
(d) a multi-task loss function with masking and coarse-level supervision to jointly optimize sound detection, azimuth, and elevation estimation, ensuring high precision in diverse acoustic conditions.
Extensive experiments demonstrate AuralNet’s superiority over the recent sector-based method \cite{yang2024deepear} across diverse conditions, including varying noise levels, reverberation, and source configurations.

\section{Proposed Model} \label{sec:methods}
We parameterize sound source DOA on a 3D unit sphere centered at the binaural microphone array, as shown in Fig.~\ref{fig:sectors}. Each source is represented by spherical coordinates $(\varphi, \theta)$, where $\varphi$ is the azimuth and $\theta$ is the elevation. 
Motivated by human auditory perception characteristics, we design a hybrid spatial encoding strategy. The azimuth is uniformly divided into $M$ sectors (each spanning $360^\circ/M$) to leverage precise binaural time-difference cues. For elevation, we adopt a symmetric partitioning scheme and divide the range $[-75^\circ, 75^\circ]$ into $N$ equal bands (each spanning $150^\circ/N$), avoiding boundaries near $\theta=0^\circ$. This design specifically addresses the perceptual ambiguity near the horizontal plane ($\theta=0^\circ$), where human hearing struggles to distinguish small elevation differences due to symmetric pinna filtering effects. As shown in Fig.~\ref{fig:sectors}(b) with $M=8$ and $N=3$, the azimuth is partitioned into eight $45^\circ$ sectors, and the elevation into three regions: $[-75^\circ, -25^\circ]$ (low), $[-25^\circ, 25^\circ]$ (central), and $[25^\circ, 75^\circ]$ (high), resulting in 24 sectors in total.
This 3D sectorization enables localization of up to $M \times N$ concurrent sound sources (assuming at most one active source per sector).

\subsection{AuralNet Overview}
Figure~\ref{fig:network} illustrates
AuralNet. Inspired by the human auditory system, AuralNet 
integrates a hierarchical coarse-to-fine multi-task learning framework, enabling simultaneous estimation of azimuth and elevation for multiple sound sources. The model is structured around three core stages: Feature Extraction, Attention-based Feature Aggregation, and Hierarchical Coarse-to-Fine Multi-Task Learning.

\subsection{Feature Extraction}
The first stage of AuralNet transforms binaural signals into time-frequency representations. To emulate the frequency response of the basilar membrane in the cochlea, we apply a Gammatone filterbank \cite{slaney1993efficient} with 64 filters that span the frequency range from $50$ to $8,000$ Hz. The signals are sampled at $16$ kHz, and each $1$-second signal from the left and right ears is framed into $50$ ms segments with $50\%$ overlap using a Hamming window. These segments are passed through the filterbank, producing log-Gammatone power spectrograms for both left and right ear signals. The output tensor shape for each ear is $(T, 64)$, where $T = 39$ represents the number of time frames.

In parallel, the GCC-PHAT algorithm is applied to extract $33$ cross-correlation (CC) coefficients from the left- and right-ear signals. The coefficients represent the interaural time differences within a range of \(\pm 1\) ms, capturing the spatial relationship between the left and right ears.



\subsection{Attention-based Feature Aggregation}
Next, we aggregate the features extracted in the previous stage using a self-attention mechanism, which allows the model to adaptively focus on the most important temporal and spectral cues within the features, facilitating more accurate SSL. To be specific, we begin by feeding the left and right Gammatone features, along with the CC features, into a dedicated attention module. In detail, the Gammatone features for each ear are first processed by individual self-attention modules to capture relevant temporal and spectral patterns in each channel. The attention mechanism is enhanced with positional encoding to maintain the temporal structure of the data, which is crucial for sequential information processing.
Meanwhie, the differences between the left and right Gammatone features are computed and passed through another attention module to capture interaural disparities, which are discriminative for SSL.

After attention-based processing, global average pooling (GAP) is applied to summarize the temporal information into a global feature vector. This pooling is performed on the outputs of the self-attention modules for the Gammatone features and the interaural feature difference, which are then concatenated with the
CC features to form the aggregated feature vector.
Finally, the concatenated feature vector is passed through a Multi-Layer Perceptron (MLP), which further refines the aggregated representation before it is sent to the hierarchical coarse-to-fine localization module.

\subsection{Hierarchical Coarse-to-Fine Multi-Task Learning}
The third stage of our model introduces a hierarchical coarse-to-fine multi-task learning framework, designed to improve SSL by jointly estimating azimuth and elevation. This approach is structured into two levels: a coarse level for preliminary predictions and a fine level for detailed refinement. The model leverages a gating mechanism and a custom weighted loss function to enhance SSL performance.

\textbf{Coarse Level:} At the coarse level, the $360^\circ$ azimuth range is divided into $M$ coarse sectors, each representing an azimuth division. The model generates a preliminary sound detection probability for each sector (a binary classification), which guides the subsequent finer-grained localization process.

\textbf{Fine Level:} Each coarse branch is further divided into $N$ fine branches, corresponding to specific elevation ranges (fine sectors). These fine branches work in tandem to refine the coarse azimuth predictions. A \textit{gating mechanism} is designed to combine coarse-level and fine-grained features:
\begin{equation}
    \boldsymbol{g} = \sigma(\mathbf{W}_g \cdot \texttt{concat}(\boldsymbol{f}_{\text{coarse}}, \boldsymbol{f}_{\text{fine}}) + \boldsymbol{b}_g),
\end{equation}
where $\boldsymbol{g}$ is the gate output, $\sigma$ the sigmoid activation function, $\mathbf{W}_g$ the learnable weight matrix, $\boldsymbol{b}_g$ the bias term, and $\boldsymbol{f}_{\text{fine}}$ and $\boldsymbol{f}_{\text{coarse}}$ represent the fine and coarse features, respectively.
The final fused feature for each fine branch is
\begin{equation}
    \boldsymbol{f}_{\text{fused}} = \boldsymbol{g} \cdot \boldsymbol{f}_{\text{fine}} + (1 - \boldsymbol{g}) \cdot \boldsymbol{f}_{\text{coarse}}.
\end{equation}
This gating mechanism allows adaptive fusion of coarse and fine features, dynamically adjusting the contribution to ensure the model focuses on the most relevant information.

Each fine branch produces three predictions for its corresponding sector: a binary classification for sound detection (indicating the presence of a source), a normalized azimuth angle, and a normalized elevation angle. The predictions from all $N$ fine branches are concatenated with the coarse detection, forming $3N + 1$ predictions for each coarse branch. Finally, the outputs from all $M$ coarse branches are concatenated along the first dimension to produce the final $(M,3N+1)$ predictions, providing 3D SSL results across all $M \times N$ sectors.

\textbf{Multi-Task Loss Optimization: } 
The network is trained with a multi-task loss function that integrates classification and regression tasks at both coarse and fine levels.
For each coarse branch $i\in\{1,2,\cdots,M\}$ and each fine sector $j\in\{1,2,\cdots,N\}$ within branch $i$, the sound detection loss is computed by using binary cross-entropy (BCE):
\begin{equation}
    \mathcal{L}_{\text{det}}^{i,j} = \texttt{BCE}(\hat{y}_{\text{det}}^{i,j}, \ y_{\text{det}}^{i,j}),
\end{equation}
where $\hat{y}_{\text{det}}^{i,j}$ is the predicted detection result, and $y_{\text{det}}^{i,j}$ is the ground-truth label.
The azimuth and elevation losses are computed using masked mean absolute error (MAE):
\begin{equation}
    \mathcal{L}_{\text{azi}}^{i,j} = \frac{1}{C_{\text{valid}}^{i,j}} \sum_{k} |\hat{y}_{\text{azi}}^{i,j,k} - y_{\text{azi}}^{i,j,k}| \cdot m_k^{i,j},
\end{equation}
\begin{equation}
    \mathcal{L}_{\text{ele}}^{i,j} = \frac{1}{C_{\text{valid}}^{i,j}} \sum_{k} |\hat{y}_{\text{ele}}^{i,j,k} - y_{\text{ele}}^{i,j,k}| \cdot m_k^{i,j},
\end{equation}
where $\hat{y}_{\text{azi}}^{i,j,k}$ and $\hat{y}_{\text{ele}}^{i,j,k}$ are the predicted azimuth and elevation for the $k$-th sample, $y_{\text{azi}}^{i,j,k}$ and $y_{\text{ele}}^{i,j,k}$ the corresponding ground-truth labels.
The term $m_k^{i,j}$ is a binary mask indicating the presence of a sound source in the $j$-th sector of the $i$-th coarse branch for the $k$-th sample, and $C_{\text{valid}}^{i,j}$ is the number of valid (active) sectors across all samples. 
This masking ensures that angular errors are only computed for sectors where a sound source is present, preventing optimization based on irrelevant regions.

To further enhance localization, we add a coarse-level supervision term using BCE loss:
\begin{equation}
    \mathcal{L}_{\text{coarse}}^i = \texttt{BCE}(\hat{y}_{\text{coarse}}^i, \ y_{\text{coarse}}^i),
\end{equation}
where $\hat{y}_{\text{coarse}}^i$ is the predicted coarse-level detection result for the $i$-th coarse branch, $y_{\text{coarse}}^i$ the corresponding ground-truth label. The total multi-task loss is a weighted sum of all losses:
\begin{equation}
    \mathcal{L} =
    \sum_{i=1}^{M} \left[ 
    \delta \, \mathcal{L}_{\text{coarse}}^i + 
    \sum_{j=1}^{N}\left( \alpha \, \mathcal{L}_{\text{det}}^{i,j} + \beta \, \mathcal{L}_{\text{azi}}^{i,j} + \gamma \, \mathcal{L}_{\text{ele}}^{i,j} \right)
    \right],
\end{equation}
where $\delta$, $\alpha$, $\beta$, and $\gamma$ are empirically determined weights, set to $0.2, 0.2, 0.5, 0.3$ in our experiments. 
This weighted masked loss function ensures that the network focuses on optimizing localization accuracy only where sound sources are present, leveraging both coarse and fine-level outputs.

\begin{table*}[t]
\setlength {\belowcaptionskip}{-0.5cm}
\caption{ Performance Comparison (Seen / Unseen Rooms) }
\label{tab: performance}
\centering
\renewcommand{\arraystretch}{0.9}
\resizebox{1\textwidth}{!}{%
\begin{tabular}{cl|cccccccccccc} 
\toprule
\multicolumn{2}{c}{Source Type}     & \multicolumn{3}{c}{1-talker}      & \multicolumn{3}{c}{2-talker}      & \multicolumn{3}{c}{3-talker} & \multirow{2}{*}{Avg.}     \\ 
\cmidrule(lr){1-2} \cmidrule(lr){3-5} \cmidrule(lr){6-8} \cmidrule(lr){9-11} 
\multicolumn{2}{c}{SNR (dB)}     & 20 & 10 & 0  & 20 & 10 & 0  & 20 & 10 & 0  \\ 
\midrule
\multicolumn{1}{c}{\multirow{4}{*}{\textbf{\begin{tabular}[c]{@{}c@{}}Sound Detection\\Accuracy (\%)\end{tabular}}}}
    & DeepEar \cite{yang2024deepear} & 98.24 / 98.33 & 98.35 / 98.34 & 97.91 / 97.81 & 93.84 / 93.73 & 93.87 / 94.03 & 93.15 / 93.13 & 89.32 / 89.33 & 89.33 / 89.21 & 88.58 / 88.84 & 93.63  \\ 
    & \textbf{AuralNet} & \textbf{99.40 / 99.39} & \textbf{99.22 / 99.23} & \textbf{98.79 / 98.71} & \textbf{95.68 / 95.52} & \textbf{95.50 / 95.46} & \textbf{94.56 / 94.51} & \textbf{90.84 / 90.69} & \textbf{90.57 / 90.43} & \textbf{89.55 / 89.65} & \textbf{94.87}\\ 
    & AuralNet-NC & 99.30 / 99.26 & 99.16 / 99.02 & 98.55 / 98.44 & 95.36 / 95.20 & 95.03 / 95.15 & 94.26 / 94.11 & 90.47 / 90.49 & 90.18 / 89.98 & 89.22 / 89.28 & 94.58\\
    & AuralNet-NH  & 99.33 / 99.28 & 99.13 / 99.07 & 98.53 / 98.54 & 95.36 / 95.17 & 95.03 / 95.05 & 94.17 / 94.02 & 90.38 / 90.40 & 90.14 / 90.12 & 89.28 / 89.35 & 94.58 \\
    & AuralNet-RL & 99.01 / 98.96 & 98.76 / 98.77 & 98.19 / 98.05 & 95.05 / 94.84 & 94.71 / 94.72 & 93.71 / 93.76 & 90.07 / 90.09 & 89.71 / 89.72 & 88.91 / 89.08 & 94.23\\
\midrule
\multicolumn{1}{c}{\multirow{4}{*}{\textbf{\begin{tabular}[c]{@{}c@{}}Sound Detection\\F1 score (\%)\end{tabular}}}}
    & DeepEar \cite{yang2024deepear} & 77.72 / 78.88 & 78.99 / 78.94 & 71.72 / 70.70 & 55.29 / 54.72 & 54.07 / 55.85 & 43.85 / 43.69 & 40.55 / 40.89 & 38.79 / 37.98 & 29.42 / 30.66 & 54.59\\
    & \textbf{AuralNet}  & \textbf{92.88 / 92.71} & \textbf{90.70 / 90.70} & \textbf{85.47 / 84.38} & \textbf{72.68 / 71.73} & \textbf{71.34 / 70.93} & \textbf{63.39 / 62.84} & \textbf{56.64 / 55.90} & \textbf{55.04 / 54.16} & \textbf{46.49 / 46.89} & \textbf{70.27}\\
    & AuralNet-NC & 91.68 / 91.23 & 90.11 / 88.35 & 82.76 / 81.32 & 70.52 / 69.90 & 68.30 / 69.30 & 61.03 / 60.20 & 54.69 / 54.90 & 52.50 / 51.56 & 44.22 / 44.57 & 68.17 \\
    & AuralNet-NH & 91.97 / 91.39 & 89.74 / 88.99 & 82.54 / 82.61 & 70.56 / 69.46 & 68.20 / 68.45 & 60.60 / 59.66 & 53.96 / 54.24 & 52.44 / 52.11 & 44.31 / 45.12 & 68.13 \\
    & AuralNet-RL & 88.07 / 87.49 & 85.06 / 85.08 & 77.78 / 76.05 & 68.03 / 66.66 & 65.35 / 65.45 & 56.18 / 56.76 & 51.07 / 51.42 & 48.35 / 48.48 & 40.66 / 42.01 & 64.44\\
\midrule
\multicolumn{1}{c}{\multirow{4}{*}{\textbf{\begin{tabular}[c]{@{}c@{}}Azimuth\\DAE ($^\circ$)\end{tabular}}}} 
     & DeepEar \cite{yang2024deepear} & 5.99 / 6.09 & 5.98 / 6.04 & 7.37 / 7.46 & 8.19 / 8.34 & 8.62 / 8.74 & 9.29 / 9.14 & 10.05 / 9.91 & 10.35 / 9.91 & 10.22 / 10.22 & 8.44\\
     & \textbf{AuralNet}  & \textbf{1.16 / 1.26} & \textbf{1.31 / 1.37} & \textbf{1.74 / 1.78} & \textbf{2.81 / 2.74} & \textbf{2.89 / 2.96} & \textbf{3.36 / 3.44} & \textbf{4.05 / 3.93} & \textbf{4.23 / 4.12} & \textbf{4.64 / 4.77} & \textbf{2.92}\\
     & AuralNet-NC & 1.71 / 1.81 & 1.83 / 1.92 & 2.24 / 2.45 & 3.21 / 3.25 & 3.32 / 3.56 & 3.96 / 3.93 & 4.68 / 4.78 & 4.77 / 4.78 & 5.36 / 5.30 & 3.49 \\
     & AuralNet-NH & 1.64 / 1.67 & 1.76 / 1.83 & 2.17 / 2.23 & 3.20 / 3.26 & 3.36 / 3.34 & 4.01 / 3.96 & 4.87 / 4.69 & 4.88 / 4.83 & 5.36 / 5.37 & 3.47 \\
     & AuralNet-RL & 18.00 / 17.95 & 17.71 / 17.83 & 18.25 / 18.22 & 18.50 / 18.72 & 18.90 / 18.92 & 18.72 / 18.91 & 19.96 / 20.08 & 20.39 / 20.23 & 19.85 / 19.89 & 18.95\\
\midrule
\multicolumn{1}{c}{\multirow{4}{*}{\textbf{\begin{tabular}[c]{@{}c@{}}Elevation\\DAE ($^\circ$) \end{tabular}}}}    
     & DeepEar \cite{yang2024deepear} & 7.33 / 7.20 & 7.65 / 7.61 & 9.12 / 9.03 & 10.06 / 9.86 & 10.15 / 10.08 & 10.52 / 10.47 & 10.88 / 10.83 & 11.17 / 10.78 & 11.36 / 11.00 & 9.73\\
     & \textbf{AuralNet} & \textbf{1.68 / 1.81} & \textbf{1.80 / 1.94} & \textbf{2.36 / 2.52} & \textbf{3.86 / 3.85} & \textbf{4.04 / 4.07} & \textbf{4.59 / 4.75} & \textbf{5.35 / 5.63} & \textbf{5.67 / 5.53} & \textbf{6.11 / 6.24} & \textbf{3.99}\\
     & AuralNet-NC & 2.38 / 2.43 & 2.47 / 2.64 & 3.12 / 3.20 & 4.75 / 4.59 & 4.73 / 4.84 & 5.23 / 5.55 & 6.14 / 6.27 & 6.45 / 6.23 & 6.75 / 7.02 & 4.71 \\
     & AuralNet-NH & 2.15 / 2.21 & 2.27 / 2.36 & 2.89 / 3.10 & 4.42 / 4.52 & 4.69 / 4.68 & 5.23 / 5.42 & 6.11 / 6.22 & 6.22 / 6.09 & 6.70 / 7.03 & 4.57 \\
     & AuralNet-RL & 19.00 / 18.96 & 18.92 / 18.74 & 19.05 / 19.32 & 19.92 / 20.26 & 19.52 / 20.23 & 19.66 / 19.53 & 20.36 / 20.63 & 20.21 / 19.71 & 20.09 / 19.76 & 19.66\\
\bottomrule
\end{tabular}%
}
\end{table*}

\section{Experimental Setup}
\subsection{Dataset}
Clean binaural signals were synthesized by convolving anechoic TIMIT speech \cite{Garofolo1993} with head-related transfer functions (HRTFs) from the ITA database \cite{Bomhardt2017}, using subject MRT05. The DOA space spanned $0^\circ$ to $360^\circ$ in azimuth and $-65^\circ$ to $75^\circ$ in elevation, both at $5^\circ$ resolution, yielding $2,088$ directions. The 3D space was divided into $M=8$ coarse azimuthal sectors ($45^\circ$ each), each further split into $N=3$ fine sectors. Elevation coverage varied across low ($81$), central ($90$), and high ($90$) regions. Sources were placed $1$ m from the microphone center, with a $0.14$ m inter-microphone distance. To simulate reverberant conditions, room impulse responses (RIRs) were generated using the image method \cite{allen1979image} and convolved with the clean binaural signals. Diffuse noise from the PCAFETER subset of the DEMAND dataset \cite{Thiemann2013} was added to simulate a noisy, reverberant, cafeteria-like environment at SNRs of $20$, $10$, and $0$ dB.

The training set included three configurations: \textit{1-talker} ($25,056$ samples), \textit{2-talker} ($43,848$ samples), and \textit{3-talker} ($29,232$ samples). Samples were generated following the above procedure, with multi-source cases enforcing a minimum azimuthal separation of $45^\circ$. Random room layouts were used, simulating small ($[5,5,3]$ m / $[6,6,4]$ m, $T_{60}=0.3/0.4$ s), medium ($[10,8,4]$ m / $[12,10,5]$ m, $T_{60}=0.5/0.6$ s), and large rooms ($[15,10,5]$ m / $[13,9,5]$ m, $T_{60}=0.7/0.8$ s). Validation used unseen rooms and speech, while testing covered both seen and unseen configurations across all source/SNR conditions. In total, the training set comprised $98,136$ samples across the three configurations. The validation and test sets contained $2,088$ single-source and $2,800$ multi-source samples, respectively.
The Adam optimizer (learning rate 0.001, batch size 200) was used, and training was stopped after 100 epochs with early stopping if no improvement was observed after 10 epochs on validation loss. 

\subsection{Evaluation Metrics}

We evaluate AuralNet using three metrics: Detection Accuracy, F1 Score, and Detection-aware Angular Error.

\textbf{Detection Accuracy} measures the binary detection accuracy of the fine branch's sound detection output.

\textbf{F1 Score} is the harmonic mean of precision and recall, measuring the model's ability to detect sound sources while minimizing false positives and negatives.

\textbf{Detection-aware Angular Error (DAE)} evaluates the average angular error for correctly detected sectors, calculated as:
\begin{equation}
    \text{DAE} = \frac{1}{N_{\text{valid}}} \sum_{n=1}^{N_{\text{valid}}} \left( \lvert \hat{\varphi}_n - \varphi_n \rvert + \lvert \hat{\theta}_n - \theta_n \rvert \right),
\end{equation}
where $\hat{\varphi}_n$ and $\hat{\theta}_n$ are the predicted azimuth and elevation (denormalized to degrees), and $\varphi_n$ and $\theta_n$ are the
ground-truth values.
$N_{\text{valid}}$ is the number of valid detections, computed as
\begin{equation}
    N_{\text{valid}} = \sum_{k=1}^{M \cdot N} \mathbb{I} \ (\hat{y}_{\text{det}}^k = 1 \, \text{ and } \, y_{\text{det}}^k = 1),
\end{equation}
where $M \cdot N$ is the total number of fine sectors, $\hat{y}_{\text{det}}^k$ is the detection result (1 for detection, 0 for no detection), and $y_{\text{det}}^k$ is the ground-truth label. 


\section{Evaluation Results and Discussion}
We evaluated AuralNet against the recent sector-based multi-BSSL system, DeepEar \cite{yang2024deepear}. To ensure fair comparison, we extend DeepEar by adapting its 2D sector division into a 24-sector 3D framework and replacing its distance estimation sub-network with our elevation estimator, resulting in a model size of $1.96$ M parameters.
To assess the effectiveness of main components in AuralNet, we compared the performance of AuralNet ($1.11$ M parameters) with its three variants:
\begin{itemize}
    \item \textbf{AuralNet-NoCoarse (AuralNet-NC)} removes the coarse detection task, bypassing the coarse output and gating mechanism, resulting in $1.00$ M parameters. 
    \item \textbf{AuralNet-NonHierarchical (AuralNet-NH)} builds upon AuralNet-NC and adopts a non-hierarchical structure with $24$ identical sub-networks for output prediction, increasing the model size to $1.53$ M.
    \item \textbf{AuralNet-RegularLoss (AuralNet-RL)} uses the standard loss function, with BCE for both coarse and fine sound detection and MAE for azimuth and elevation prediction, maintaining the same architecture as AuralNet ($1.11$ M).
\end{itemize}

The results summarized in Table \ref{tab: performance} indicate that AuralNet consistently outperforms DeepEar across all tasks and noisy-reverberant conditions, with significant improvements in F1 score and DAE in both azimuth and elevation estimation.
AuralNet achieved the highest detection accuracy and F1 score across all SNRs and room configurations. At $20$ dB SNR, AuralNet’s F1 score for 1-talker in unseen rooms was $92.71\%$, outperforming DeepEar's $78.88\%$ by nearly $14\%$. The F1 score was lower than accuracy due to class imbalance, with silent sectors dominating the dataset. Despite this, the significant improvement in F1 score highlights AuralNet’s superior ability to handle complex localization tasks. 

In terms of DOA estimation, AuralNet demonstrated a substantial advantage in both azimuth and elevation DAE acorss all scenarios. AuralNet achieved an average azimuth DAE of $2.92^\circ$, a remarkable improvement over DeepEar’s $8.44^\circ$. Similarly, AuralNet’s average elevation DAE was $3.99^\circ$, significantly better than DeepEar’s $9.73^\circ$. These results show that AuralNet’s architecture enables more accurate DOA estimation, demonstrating its potential for real-world BSSL tasks.



The comparison of AuralNet with three variants showed important performance trade-offs. AuralNet-NC, which omitted the coarse detection task, exhibited slightly lower performance, particularly in sound detection and F1 score. This highlights the crucial role of the coarse detection task and its supervisory effect through the coarse loss, which is beneficial for the subsequent fine tasks. AuralNet-NH, despite having a larger model size, showed similar performance to AuralNet-NC in all tasks, underscoring the efficiency of the hierarchical structure. This suggests that the hierarchical approach can reduce model size and redundancy without sacrificing performance. Lastly, AuralNet-RL, which used a standard loss function, performed the worst, especially in DOA estimation. This stark performance drop highlights the effectiveness of AuralNet’s masked loss function in improving DOA accuracy.


Overall, AuralNet demonstrated significant improvements over DeepEar in both sound detection and localization tasks. Its architecture, including the coarse detection task, hierarchical structure, and custom loss function, contributed to its superior performance, making it a promising solution for real-world multi-BSSL scenarios.


\section{Conclusions}
We have proposed AuralNet, a novel hierarchical attention-based model for 3D 
BSSL of multiple concurrent speakers. Unlike existing methods limited to 2D or single-source localization, AuralNet jointly estimates azimuth and elevation in 3D, requiring no prior knowledge of the number of sources. By utilizing a coarse-to-fine multi-task learning strategy and integrating self-attention mechanisms for binaural feature aggregation, AuralNet significantly outperforms DeepEar \cite{yang2024deepear}, a recent sector-based multi-BSSL system, in noisy-reverberant environments, especially in low-SNR and multi-speaker scenarios. 
Future work will focus on improving the model's generalizability to more challenging acoustic environments and to optimize its performance for real-world embodied applications. 

\section{Acknowledgements}
This work was supported by the Science, Technology, and Innovation Commission of Shenzhen Municipality, China (Grant No. ZDSYS20220330161800001), the Shenzhen Science and Technology Program (Grant No. KQTD20221101093557010), and the Guangdong Science and Technology Program (Grant No. 2024B1212010002).


\bibliographystyle{IEEEtran}


\end{document}